# Phonon thermal conduction in novel two-dimensional materials

Xiangfan Xu[1,2,3], Jie Chen[1,2,3], Baowen Li[4]

[1] Center for Phononics and Thermal Energy Science, School of Physics Science and Engineering, Tongji University, Shanghai 200092, China
[2] China-EU Joint Lab for Nanophononics, Tongji University, Shanghai 200092, China
[3] Institute for Advanced Study, Tongji University, Shanghai 200092, China
[4] Department of Mechanical Engineering, University of Colorado, Boulder, CO 80309, USA

E- mail: xuxiangfan@tongji.edu.cn and jie@tongji.edu.cn and Baowen.Li@Colorado.edu

**Abstract**

Recently, there have been increasing interests in phonon thermal transport in low dimensional materials, due to the crucial importance for dissipating and managing heat in micro and nano electronic devices. Significant progresses have been achieved for one-dimensional (1D) systems both theoretically and experimentally. However, the study of heat conduction in two-dimensional (2D) systems is still in its infancy due to the limited availability of 2D materials and the technical challenges in fabricating suspended samples suitable for thermal measurements. In this review, we outline different experimental techniques and theoretical approaches for phonon thermal transport in 2D materials, discuss the problems and challenges in phonon thermal transport measurements and provide comparison between existing experimental data. Special focus will be given to the effects of the size, dimensionality, anisotropy and mode contributions in the novel 2D systems including graphene, boron nitride, $MoS_2$, black phosphorous, silicene etc.

**Contents**

## 1. Introduction

The thermal properties of low-dimensional systems are of interests for both fundamental researches and applications. For applications, the accumulated heat spot in high-density electronic devices has become the bottleneck for further miniaturization of the modern electronics. The heat generation of the electronic and optoelectronic devices can raise their operating temperature to the point when thermal management becomes critical, which often limits the devices performance, and in the worst case can lead to device failure. Of particular importance is to increase the thermal conductance of low dimensional materials, and to control the thermal interfacial resistance to be in a relative low level. From fundamental perspective, there are great demands to understand heat conduction in low dimensional systems. In fact the size-effect and nonlinear effect are ignored in the traditional thermal transport theory, resulting in unclear physics behind the thermal dissipation and thermal management in nano/micro scale. Fortunately, the successful exfoliation of graphene and the related 2D materials (2D) [1-10] provide perfect test platform of deep insight into transport properties of 2D phonons and their interactions [11-16].

Phonons - the quantized collective modes of crystal lattice vibrations, especially the acoustic phonons, are the main heat carriers of semiconductors and insulators. The rapid progress of nanotechnology achieved in last two decades shows that phonon thermal conduction in nanoscale, such as thin films and nanowires, dramatically differ from that in their bulk counterparts and are strongly suppressed due to the increase of phonon-boundary scatterings, changes in phonon group dispersion and phonon density of states [17-19]. These phenomena lead to the possibility of phonon control and heat management in nanoscale, and trigger the recent applications of nanomaterials in thermoelectrics and thermal insulator materials. On the other hand, phonons in quasi one-dimensional materials (1D) such as carbon nanotube and boron nitride nanotube, and 2D materials such as graphene and boron nitride, behavior differently and can conduct heat more efficiently than their bulk counterparts [20-23]. This exotic behavior, together with invalidation of Fourier's law [11, 24-31], attract world-wide interests and led to unsettled hot discussions to date.

In the last decade, various theories have been developed to investigate the underlying physical mechanism of heat transport in low-dimensional materials, such as molecular dynamics (MD) simulations [32-34], non-equilibrium Green's function (NEGF) method [35-37], and Boltzmann transport equation (BTE) [38-40]. In contrast, experimental studies on low dimensional materials, especially 2D materials, are relatively rare, due to the challenges in suspending nanomaterials suitable for thermal measurements and in measuring the temperature distribution in nano/micro scale.

In this review, we present the recent progress in understanding the phonon thermal transport in 2D materials both experimentally and theoretically. In Sec. 2, we discuss the experimental setups for thermal transport measurements, together with the related drawbacks and challenges. In Sec. 3, we show the experimental results on the novel 2D materials, including graphene, boron nitride, $MoS_2$ and black phosphorous. We introduce the novel phonon thermal transport in 2D materials, such as length dependence, thickness dependence and anisotropic effect in Sec. 4-6. We also briefly review the various theoretical approaches for investigating thermal transport in Sec. 7. In the last section, Sec.8, we give conclusions and outlooks.

It is important to note that the selected 2D materials belong to a huge 2D material family [8]. However, for most of materials, phonon thermal conduction remains untouched. The combination of experimental and theoretical effects will help in establishing a clearer picture of the phonon thermal conduction in 2D materials and shed the light for utilizing 2D materials as potential materials for thermal managements, thermoelectrics and information carriers [41].

Due to the length limit, we only address the fundamental phonon thermal properties of 2D materials. We note that there are plenty of studies and reviews on different aspects of thermal properties in nanostructured materials. For comprehensive reviews please refer to references [41-46]; for progress of experimental studies in nanoscale thermal transports, please refer to references [47-50]. There are also reviews on anomalous and exotic thermal transports in low dimensional materials [14, 16, 51].

## 2. Experimental setups of phonon thermal transport in 2D materials

The success in measuring thermal conductivity in low dimensional materials has helped to understand 2D phonons from both fundamental and application point of views [38, 52-59]. Of particular interests are the thermal conductance of 2D materials and the interfacial heat transfer between 2D materials and the substrates [60-62], which play a critical role in 2D field-effect transistor performance and current saturation.

## 2.1 Experimental setups

The modern nano-fabrication technologies enable us to heat nanoscale materials and to measure temperature gradient at the same time, although still with great challenges. Various techniques have been invented to measure the intrinsic thermal conductivity of low-dimensional materials, e.g. confocal micro-Raman method [23], thermal bridge method with prepatterned built-in microstructures [22, 63-65], 3ω method [66, 67], time-domain thermoreflectance (TDTR) method [68], scanning thermal microscope (SThM) [69, 70] and thermal flash method [71]. In this section, we briefly review these technologies. Special attention will be paid to two popular methods: confocal micro-Raman method and thermal bridge method.

**Confocal micro-Raman method**

Raman method is the most popular and approachable technique for measuring thermal conductivity of low-dimensional materials. In this section, for convenience, we will briefly discuss the details of measuring thermal conductivity in suspended graphene.

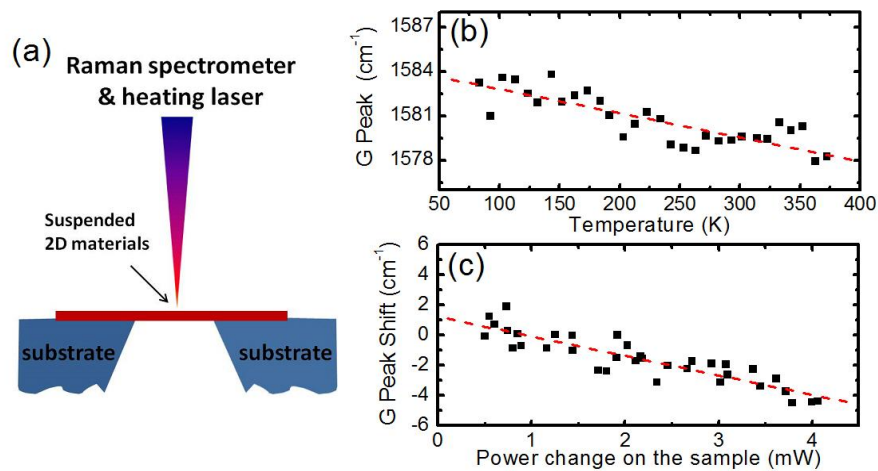

**Figure 1**. (a)Experimental setup of confocal micro-Raman method. (b)Experimental data for Raman G-peak shift with respect to temperature. Reprinted with permission from[72]. Copyright 2007 American Chemical Society. (c) Experimental data for Raman G-peak shift versus laser power. Reprinted with permission from[23]. Copyright 2008 American Chemical Society.

Figure 1a shows the setup of confocal micro-Raman method for measuring thermal conductivity of single layer and few layer graphene [14, 23, 52]. Graphene sample was either exfoliated through standard scotch tape method [23] or directly transferred from CVD-graphene/copper onto trenches or corbino holes [52]. The suspended part of graphene cross trenches and corbino holes are the objects need to be measured, and the supporting part can be regarded as contact and heat sink. The sample was kept in vacuum to reduce heat leakage through air. A high energy laser beam with spot size of 0.5 μm to 1μm was focused on the center of sample. The laser beam has two purposes: heat the sample and detect the temperature rise by Raman spectroscope.

For graphene suspended cross trench, one can assume that heat spreads from center to two opposite directions when the width of trench is much larger than laser spot size. In this case, thermal conductivity $\kappa$ of suspended graphene can be obtained by [23]:

$$\kappa = \left(\frac{L}{2S}\right)\left(\frac{\Delta P}{\Delta T}\right), \tag{1}$$

where $L$ is distance from center of laser spot to intersection of graphene and substrate, i.e. half of sample length, $T$ is the absolute temperature, $S = h \times w$ is the cross section area of graphene perpendicular to the spreading direction of heat, $w$ is the width of sample, $h$ is thickness of single layer graphene which is measured to be around 0.34nm. $\Delta P$ is laser power absorbed by graphene, which can be either directly measured by laser power meter, or assumed 2.3% -3.6% absorption for one layer of graphene.

For graphene sample with corbino geometry or laser spot size is comparable with the width of trench, thermal conductivity of suspended graphene can be obtained by [23]:

$$\kappa = \left(\frac{1}{2\pi h}\right)\left(\frac{\Delta P}{\Delta T}\right). \tag{2}$$

The temperature rise $\Delta T$ in center of graphene can be detected by measuring the fingerprint of G-peak with temperature from Raman spectroscope, which was firstly studied by Calizo *et al*. [72]. Figure 1b represents the experimental data of Raman G-peak shift as function of temperature. The G-peak shifts linearly with temperature changes by $\omega = \omega_0 + \chi T$, where $\omega$ is Raman G-peak shift frequency, $\omega_0$ is Raman G-peak shift frequency at absolute zero Kelvin, $\chi$ is temperature coefficient which is measured to be $\chi = -(1.62 \pm 0.20) \times 10^{-2}$ cm$^{-1}$/K for suspended single layer graphene.

During Raman measurement, laser power is increased gradually, resulting in a G-peak shift with laser power (Figure 1c) [23]. Therefore, formula (1) and (2) can be modified into following formula and thermal conductivity can be given by measuring $\chi$ and $\delta\omega/\delta P$:

$$\kappa = \chi\left(\frac{L}{2hW}\right)\left(\frac{\delta\omega}{\delta P}\right)^{-1}, \tag{3}$$

$$\text{and} \quad \kappa = \chi\left(\frac{1}{2h\pi}\right)\left(\frac{\delta\omega}{\delta P}\right)^{-1}. \tag{4}$$

Based on the discussions above, confocal micro-Raman method appears to be one of the most commonly used technique due to its easy access and has been very successfully in measuring room temperature thermal conductivity of 2D materials [23, 56], dimensional crossover effect [12], isotopic effect [13] and anisotropic effect etc [73].

**Thermal bridge method**

Before the invention of thermal bridge method, thermal conductivity of multi-walled carbon nanotube was measured in the form of bundles using standard steady-state method and its value was determined to be extremely low due to the scatterings in the barriers between tubes[74, 75]. In 2001, P. Kim *et al.* introduced the thermal bridge method by integrating complex electron beam lithography and nano-manipulation to measure thermal conductivity of individual multi-walled carbon nanotube [22]. A suspended microelectromechanical system (MEMS) device and nano-manipulation system were used to suspend low-dimensional materials and detect temperature changes in micro/nano scale.

The MEMS devices are mass-fabricated by a standard wafer-stage nanofabricating process, as shown in Figure 2. A 300nm- to 500nm- thick $SiN_x$ film is fabricated by low strain PECVD method (Figure 2a), followed by standard electron beam lithography (or deep-UV lithography) and e-beam deposition (Figure 2b). Two Pt/Cr coils, acting as heater and temperature sensor, each connected by six beams with 500 μm in length, are fabricated during this process. At a second step of lithography, photoresist is patterned to cover the Pt electrodes (Figure 2c), followed by the reactive ion etching (RIE) to etch away $SiN_x$ film which is not protected by photoresist (Figure 2d). At final step, photoresist is removed (Figure 2e) and the whole MEMS device is dipped in KOH or TMAH solution for suspension (Figure 2f).

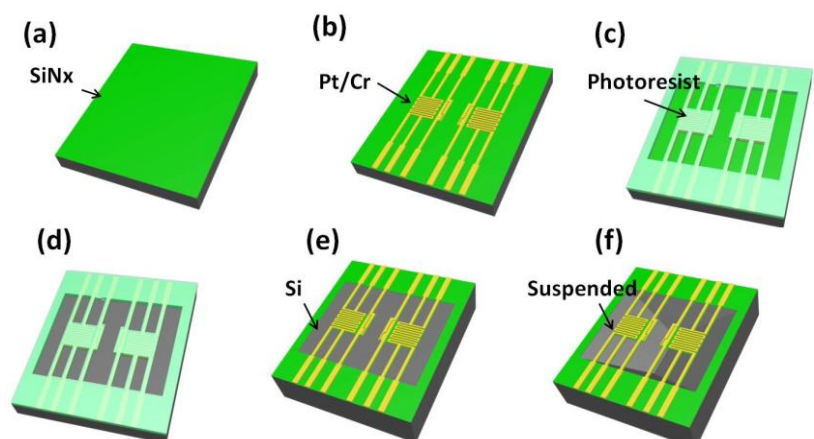

**Figure 2**. Fabrication processes of MEMS devices.

The suspended sample, either 1D materials or 2D materials, provides a thermal path between the two $SiN_x$ membranes (Heater and Sensor) that are otherwise thermally and electrically isolated from each other (Figure 3a). A μA-DC current combined with an AC current (100-200 nA) is applied to the heater resistor ($R_h$,). The DC current ($I_{DC}$) is used to apply Joule heat in $R_h$ and to increase its temperature ($T_h$) from the environment temperature, $T_0$. The AC current is used to measure the resistance of $R_h$, corresponding to $T_h$ as Pt metal is a good thermometer. The Joule heat with heating power of $Q_h = I_{DC}^2 R_h$ in Heater gradually dissipates through the six Pt/SiN$_x$ beams and the sample connecting them, which rises the temperature ($T_s$) in the sensor resistor ($R_s$). At the meaning time, the DC current also heats the two Pt/SiN$_x$ beams with a Joule heat power of $2Q_L$, half of which conducts to heater and the other half dissipates through the Pt/SiN$_x$ beams, making the total Joule heat power on heater $Q_h + Q_L$. In the steady state, thermal resistance circuit (see Figure 3b) of MEMS device can be explained by

$$Q = Q_1 + Q_2 = Q_h + Q_L, \tag{5}$$

$$Q_1 = G_b \times \Delta T_h, \tag{6}$$

$$Q_2 = G_s(\Delta T_h - \Delta T_s) = G_b \times \Delta T_s, \tag{7}$$

and the thermal conductivity can be obtained by:

$$\kappa = G_s \frac{L}{S} = \frac{L}{R \times S}, \tag{8}$$

where $G_b$ is thermal conductance of six Pt/SiN$_x$ beams, $G_s$ is the thermal conductance of sample, $L$ is sample length, $S$ is the cross section area of sample, $R$ the total thermal resistance.

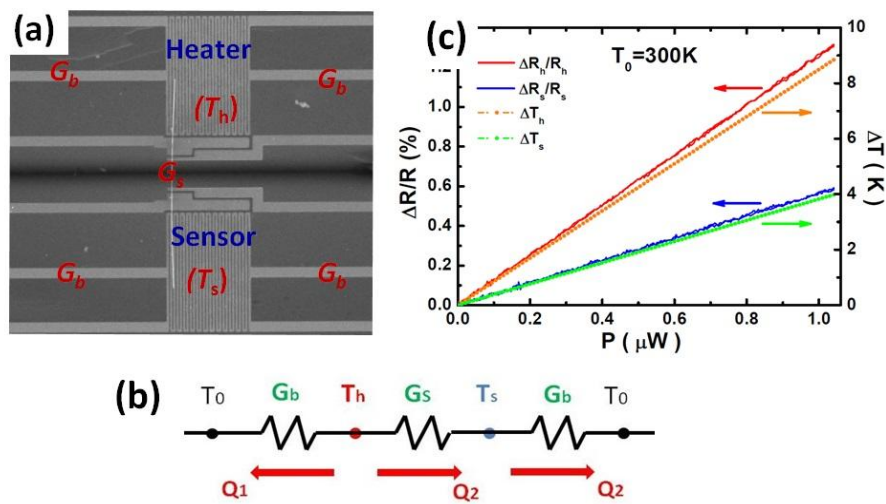

**Figure 3**. Thermal resistance circuit and measurement details for MEMS device. (a) SEM image of an insulating nanowire suspended by MEMS for thermal measurement. This measurement can give information on thermal conductivity, thermopower and electrical resistance in a single device. (b) Schematics of thermal flow circuit. (c) Resistance change and the corresponding temperature change in Heater/Sensor.

It worth noting that during the measurement, temperature change in each membrane as the function of applied Joule heat power is kept in the linear range (Figure 3c). The increases in temperature, $\Delta T_h$ and $\Delta T_s$, are controlled to be in few Kelvin (Figure 3c) to minimize thermal radiation between the two membranes. All measurements should be performed under vacuum conditions better than $1\times10^{-5}$ mbar and device should be mounted with special care such that thermal radiation to background can be eliminated.

**Other measurement techniques**

Various approaches have been utilized to measure the thermal transport properties of low dimensional materials, e.g. TDTR method [68], 3ω method [66, 67], SThM method [69, 70] and thermal flash method[71]. However, these methods can only be used to detect the thermal interfacial resistance, out-of-plane thermal conductivity or thermal conductivity in bulk materials. Although 3ω method can manage to measure thermal conductivity of nanowire [76], there is no report on the in-plane thermal conductivity of 2D materials measured by 3ω method. Therefore, we will not give detailed discussion on these techniques. For recent progress of nanotechnology on thermal transport measurements, please refer to reference [50].

**2.2** Problems in the existing experimental setups

Despite the recent advances in developing new technologies and new equipments for measuring thermal transports in low dimensions, there are plenty of problems and challenges in controlling heat flow and detecting temperature in micro/nano scale.

The commonly used confocal micro-Raman method can only detect temperature at room temperature and above, and its measurement uncertainty is known to be as high as 30% - 50% [14, 55]. The primary difficulty lies in the fact that Raman peak shifts weakly with temperature changes. For example, first-order temperature coefficient of the $E^1_{2g}$ or $A_g$ modes in $MoS_2$ is $-1.32\times10^{-2}$ cm$^{-1}$/K and $-1.23\times10^{-2}$ cm$^{-1}$/K, respectively [77]. Considering the Raman measurement accuracy of around 0.1 cm$^{-1}$ to 0.5 cm$^{-1}$, the measured temperature sensitivity is about ~8K to ~40K, in sharp contrast with that in thermal bridge measurement (~100 mK, which can be further improved by modifications [64, 65]). It is worth noting that with the help of analytical or numerical treatment of temperature distribution and heat diffusion, Stoib *et al*. has systematically discussed the application range, such as the size of sample, of Raman method, and managed to apply this method into suspended films and bulk single crystals with a much higher measurement sensitivity of ~ 0.1 Wm$^{-1}$K$^{-1}$ [78].

The other problem in Raman measurement, similar to all other thermal measurements, is the thermal contact resistance, which unavoidably contributes to the total measured thermal resistance, i.e. $R_{total} = R_{int} + 2R_c$, where $R_{total}$ is the total measured thermal resistance, $R_{int}$ is the intrinsic thermal resistance of sample, $R_c$ is the thermal contact resistance. As we know, in electrical transport measurements, the four probe method is used to reduce the effect from the contact resistance. Analogous to electrical measurement, two-laser Raman thermometry method was invented to measure intrinsic thermal conduction in silicon film [79]. In this method, Reparaz *et al*. used one laser beam focused on the center of sample to create a thermal gradient, and mapped the temperature distribution using another laser from Raman spectroscope (Figure 4a). Their results demonstrated the potential of this new contactless method for quantitative determination of thermal conductivity, however no such experiment on 2D materials has been reported yet, probably due to the inferior spatial resolution of laser spot (0.5μm to 1μm) when compared to the sample size (few micrometers to tens of micrometers) of suspended 2D materials.

On the other hand, the main challenges of thermal bridge method lie in the uncertainty contribution from the contact resistance at the two ends of samples, similar to the situation in Raman measurement. To reduce the influence from the contact resistance, Xu *et al*. deposit Cr/Au bars on the two end of graphene to create additional channel for heat to flow (Figure 4b) [11]. However, the systematical measurement on length-dependent thermal resistance of the suspended graphene shows that the effect from contact can't be eliminated to below the negligible level.

Measuring the thermal contact resistance directly is still challenging. Nevertheless, there have been many efforts trying to overcome this challenge, e.g. Wang *et al*. and Liu *et al*. have invented a brand-new method, noncontact self-heating technique, to measure the extrinsic thermal contact resistance directly (Figure 4d&e) [59, 80]. This method is modified from standard thermal bridge method by utilizing electron beam as heating source. As shown in Figure 4d, when focused electron beam scanning across the SiGe/NiSiGe bamboo structure, a clear linear relation between $R_i$ and $x$ can be observed, which indicates that the thermal contact resistance has been get rid of in this technique [80]. This technique provides direct measurement of intrinsic thermal resistance of nanowires and supported graphene, however extension of such technique to suspended 2D materials has not been reported.

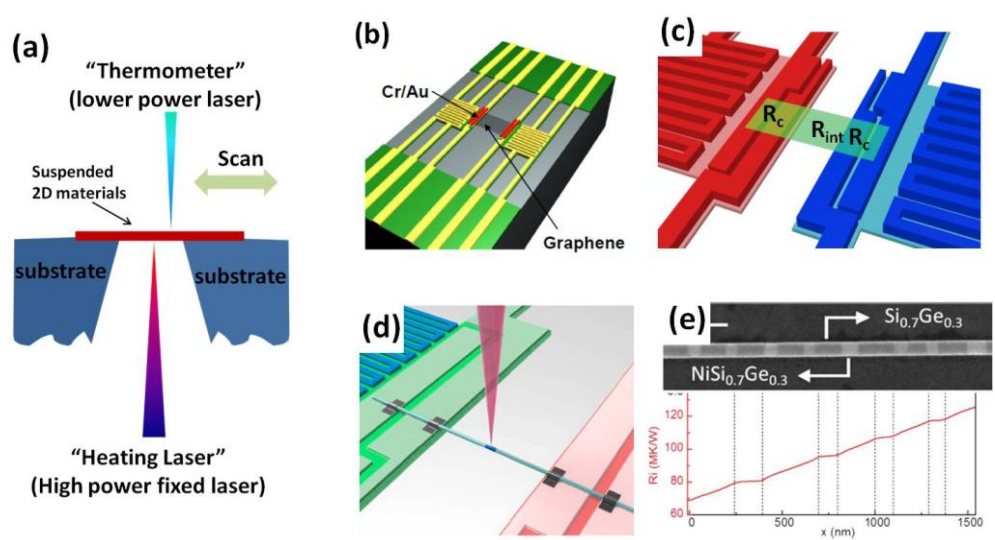

**Figure 4**. Measuring the intrinsic thermal conduction of nanostructures. (a) Experimental setup for two-laser Raman thermometry method. (b) Schematics of a graphene sheet clamped by Cr/Au bars.

(c) Schematics of heat flow in thermal bridge method. (d) Experimental setup for noncontact self-heating technique, which is modified from the standard thermal bridge method. (e) Bamboo-structured nanowire (upper panel) and the measured thermal resistance ($R_i$) versus sample length (*x*) (lower panel). Reprinted with permission from [80]. Copyright 2014 American Chemical Society.

Due to the challenges discussed above, and in additional to the effect from defects[81, 82], rough edges, thickness non-uniformity and lateral sizes (length and width), the measured thermal conductivity in 2D materials differ from groups to groups and its value scatters in several folds, leaving the intrinsic thermal properties of 2D materials unsolved with hot debates and arguments to date. Furthermore, complex nanofabrication process is applied to transfer 2D materials to prepatterned nanostructures suitable for thermal measurement. Polymer residues, cracks and ripples can be introduced into the suspended 2D materials. As discussed later on this manuscript, the absorbed particles, especially the polymer residues can reduce the thermal conductivity of suspended 2D materials to be in a level comparable to that in supported ones, which obscures the intrinsic thermal conduction of the 2D phonons and results in invisibility of novel phonons thermal transport in 2D materials, e.g. size dependence, mode's contributions etc.

## 3. Phonon thermal transport in 2D materials

### 3.1 Graphene

**Thermal conductivity near room temperature**

Recent advances in nano/micro scale fabrications and measurement techniques mentioned above have accelerated studies of thermal transports in low-dimensional materials. However, there was no progress on phonon thermal conduction of 2D materials until four years after graphene was exfoliated. The first experiment was carried out on graphene by A. Banlandin *et al*. in 2008 using confocal micro-Raman method [23]. The experiment was carried out on suspended single layer graphene which was placed in air. The room temperature thermal conductivity was measured to be as high as ~4800 $Wm^{-1}K^{-1}$ to 5300 $Wm^{-1}K^{-1}$, which is two to three times larger than that in graphite and comparable to that in diamond. It is also found that the main carriers in suspended graphene are acoustic phonons, which carry at least 99% of the heat at room temperature with the mean free path around ~775 nm [54]. However, there are hot debates on this superior value of thermal conductivity, as the authors assumed 11% to 12% absorption of laser power from graphene, which is much higher than the theoretical prediction and experimental observations [52]. When using 2.3% to 3.3% absorption, the final thermal conductivity should be around 920 $Wm^{-1}K^{-1}$ to 1600 $Wm^{-1}K^{-1}$, two to five times smaller than the original result.

Several studies utilizing confocal micro-Raman method found similar high thermal conductivity in graphene. To reduce measurement uncertainty, a laser power meter was used to directly measure the power absorbed by graphene. The obtained thermal conductivity is (2500+1100/-1050) $Wm^{-1}K^{-1}$ near 350K and reduces to (1400+500/-480) $Wm^{-1}K^{-1}$ at 500K, due to the phonon-phonon scattering [56]. Another experiment carried out by the same authors reported thermal conductivity of CVD graphene to be (2.6±0.9) to (3.1±1.0)×10$^3$ $Wm^{-1}K^{-1}$ near 350K by placing samples into vacuum [52]. They also determined the heat transfer coefficient for air and $CO_2$ to be (2.9+5.1/-2.9) and (1.5+4.2/-1.5)×10$^4$ $Wm^{-2}K^{-1}$ respectively when graphene was heated to about 510K [52].

Similar to the isotopic effect in carbon nanotube and silicon nanowires, isotopic doping introduced reduction on thermal conductivity in graphene was also observed [13]. By introducing $^{13}CH4$ into $^{12}CH4$ during CVD growth, Chen *et al*. found thermal conductivity in 0.01% $^{13}C$ graphene (regards as isotopic pure graphene) reduces about 30% to 40% when increasing $^{13}C$ concentration to 1.1% (regards as natural graphene). This is understandable that within the assumption of single relaxation time mode, phonon relaxation time can be expressed by: $\tau^{-1} = \tau_g^{-1} + \tau_p^{-1} + \tau_u^{-1}$, where $\tau_g$ is scattered by boundaries, $\tau_u$ is from U-scattering. $\tau_p$ is from defect and doping, which is related to the changes of atomic mass, e.g. $\tau_p \sim (\Delta M/M)^2$. Here, $\Delta M/M$ is the atomic mass changes from isotopic doping.

Several independent groups also found that thermal conductivity in graphene is smaller than that in bulk graphite using confocal micro-Raman method. Lee *et al*. [55] found its value to be around ~1800 $Wm^{-1}K^{-1}$ at 325K and Faugeras *et al*. [53] determined its value to be ~632 $Wm^{-1}K^{-1}$ at 350K. However, this discrepancy is due to the inaccuracy in determining the laser absorption coefficient and the temperature rise. By using the modified parameters, the measured thermal conductivity will be ~2700 $Wm^{-1}K^{-1}$ at 325K in reference [55] and ~632 $Wm^{-1}K^{-1}$ at 660K in reference [53], which is consistent with the previous studies.

Alternatively, Xu *et al*. managed to suspend single layer graphene across prepatterned MEMS and found that the room temperature thermal conductivity is around ~1605 $Wm^{-1}K^{-1}$ to ~1878 $Wm^{-1}K^{-1}$ of the sample with 9 μm in length and 1.5μm in width by thermal bridge method [11]. It is argued this relatively lower value due to difference in sample size, which will be discussed in details later on this manuscript.

**Table 1**. Thermal conductivity of graphene from different independent groups.

| Graphene source | CVD | CVD | Exfoliated | Exfoliated |
|---|---|---|---|---|
| κ near RT (W/mK) | ~1605-~1878 | ~2500-~3100 | ~1800-~5300 | ~650 |
| Layers | Monolayer | Monolayer | Monolayer | Bilayer |

| Method | 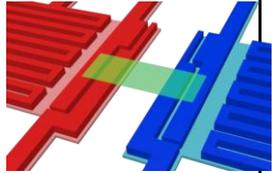 Thermal bridge | 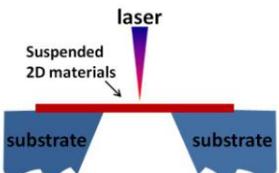 Raman | | 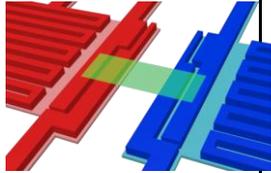 Thermal bridge |
|---|---|---|---|---|
| Geometry | rectangle | corbino | corbino & rectangle | rectangle |
| Temperature | 300K | ~350K | ~350K | 300K |
| Reference | [11] | [13, 52, 56] | [12, 23, 38, 54, 55] | [58] |

On the other hand, thermal conductivity of supported single layer graphene is much smaller due to the flexural acoustic (ZA) phonons suppressed by substrate or adhesive materials on sample surfaces [57, 83]. Experiment carried out on graphene/$SiO_2$ reveals thermal conductivity in supported graphene is ~600 $Wm^{-1}K^{-1}$. It is proposed that ZA phonons contribute around 77% of heat conduction at room temperature, which will be totally suppressed by substrate [83] or organic residues on the surface [58]. By solving Boltzmann transport equation, the intrinsic thermal conductivity of single layer graphene is calculated to be around 2600 $Wm^{-1}K^{-1}$ in reference [83].

**Thermal conductivity at low temperature**

Despite the several-folds variation of measured room temperature thermal conductivity of suspended single layer graphene, it is accepted that its value increases with decrease temperature due to the reduction of U-scattering and reaches a peak below 200K. At lower temperature, phonons will move without scattering in clean infinite graphene sheet. This is also called ballistic phonon conductance, and therefore thermal conductance per unit cross section area $\sigma/A$ is more intrinsic [15, 21, 84]. Generally speaking, $\sigma/A$ has been expected to follow $\sigma/A \approx \left[\frac{1}{4.4\times10^5 T^{1.68}} + 1/(1.2\times10^{10})\right]^{-1}$

$Wm^{-2}K^{-1}$ in the ballistic regime, as shown by the black dashed curve in Figure 5. Two independent groups, Xu *et al*. [11] and Bae *et al*. [15] observed a quasi-ballistic phonon conduction in suspended and supported submicron graphene, respectively. The experimentally measured values are within 40% of the predicted ballistic thermal conductance at $T = 30$ K [11] (Figure 5). This result is consist with their MD simulation that thermal conductance remains constant for length (between the hot and the cold reservoir) up to ~ 80 nm at 300K.

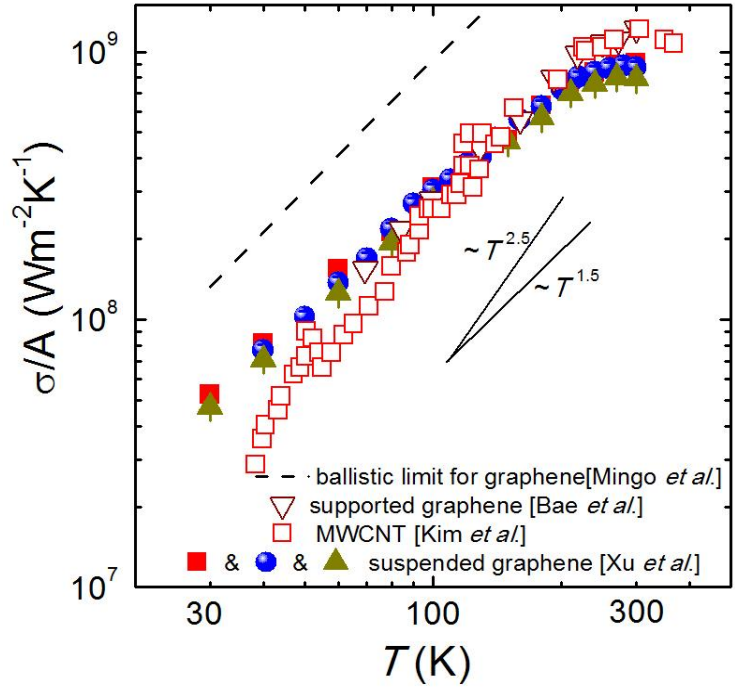

**Figure 5**. Experimental observation of ballistic thermal conductance at low temperature in submicron suspended graphene (solid squares, solid circles and solid triangles) and supported graphene (hollow triangles, data from reference[15]). Experimental data from single MWCNT is shown for comparison[22]. Reproduced with permission from [11]. Copyright 2014 Nature Publishing Group.

An unsettled debate of intrinsic thermal conduction of single layer graphene is related to which phonon mode accounting for the superior thermal conductivity. Acoustic vibrations in a 2D graphene lattice are composed of two types of phonons: in-plane phonons (TA and LA phonons) with a linear dispersion, and out-of-plane phonons (ZA phonons or flexural acoustic phonons) with a quadratic dispersion. Based on Klemens approximation [85], Nika *et al.* claimed that the contribution from ZA phonons is neglected due to its large Grüneisen parameter and small group velocity [86]. Meanwhile, Mingo *et al.* have argued that the ZA phonons carry most of the heat in single layer graphene [39]. At low temperatures, the out-of-plane ZA modes are predicted to lead to a ~$T^{1.5}$ behavior of the thermal conductivity, while the in-plane LA and TA phonons lead to a ~$T^2$ behavior. Xu *et al*. firstly carried out experimental measurement of suspended single layer graphene at low temperature and found its thermal conductance follows ~ $T^{1.5}$ to ~ $T^{1.6}$ in sample with length of 300nm and found its room temperature thermal conductivity reaches ~225 wm$^{-1}$K$^{-1}$ [87]. Although Petters *et al*. argued that this measured value is considerably lower than the theoretical prediction of the ZA contribution and attributed the reduction to the scatterings from the organic residue on the surfaces [58], it is important to note that in the ballistic regime thermal conductivity loses its role in describing the thermal conduction and $\sigma/A$ is more intrinsic [84]. The observation high value on $\sigma/A$ which approaches the quasi-ballistic phonon conduction (see Figure 5), despite the low value in thermal conductivity due to the relatively small sample size, probably indicates that the experimental data observed in submicron suspended graphene is intrinsic.

Alternatively, several indirect observations have demonstrated that the dominating carriers in intrinsic single layer graphene are ZA phonons. Combining the Boltzmann transport equation and experimental results from supported graphene, Seoul *et al*. claimed that ZA phonons contribute 77% and 86% of heat conduction at room temperature and at low temperature, respectively [83]. Wang *et al*. observed a reduction of 82% in thermal conductivity of suspended tri-layer graphene with gold deposition on surfaces and argued that ZA phonons are suppressed due to the scatterings with gold atoms [57].

### 3.2 Boron Nitride

Hexagonal boron nitride (*h*-BN), analogous to graphene, is stacked by one-atomic-thick layers of boron and nitride atoms with honey-cone structure [88]. Due to their geometry similarity, *h*-BN holds similar physical properties such as high temperature stability and superior thermal conductivity. Theoretical calculations has revealed that room temperature thermal conductivity of single layer *h*-BN reaches ~600 $Wm^{-1}K^{-1}$ [89-91] when considering the exact solution of Boltzmann transport equation, comparing to that of ~390 $Wm^{-1}K^{-1}$ in high-quality bulk *h*-BN [92-94].

Nevertheless, seldom experiments are reported on thermal conductivity of *h*-BN, especially on single layer *h*-BN. The confocal micro-Raman method is not so appreciated in *h*-BN due to its weak intensity of Raman peaks. Zhou *et al*. took the first shot on the Raman method and found first-order temperature coefficients for monolayer (1L), bilayer (2L) and nine-layer (9L) *h*-BN sheets to be $-(3.41 \pm 0.12) \times 10^{-2}$, $-(3.15 \pm 0.14) \times 10^{-2}$ and $-(3.78 \pm 0.16) \times 10^{-2}$ $cm^{-1}K^{-1}$, respectively [95]. The room-temperature thermal conductivity of 9L *h*-BN sheets was found to be ~ 227 $Wm^{-1}K^{-1}$ to 280 $Wm^{-1}K^{-1}$, which is lower than that in bulk *h*-BN (Figure 6).

Thermal conductivity of partially suspended few-layer *h*-BN was later measured by Jo *et al*. using thermal bridge method with modified built-in thermometers [96]. The highest thermal conductivity observed in 11-layer *h*-BN reaches ~360 $Wm^{-1}K^{-1}$ at room temperature, comparable to that in bulk *h*-BN (blue triangles and green squres in Figure 6). However, in contract to the thickness dependent in few-layer graphene [12], thermal conductivity in 5-layer *h*-BN is much smaller than that in bulk *h*-BN and reaches a low value of ~250 $Wm^{-1}K^{-1}$. Authors attributed this anomalous to the scattering of low frequency phonons by polymer residue, which has been observed in bilayer graphene [58]. It is worth noting that by measuring several samples with similar thickness but various lengths, authors managed to measure the contact thermal resistance between *h*-BN and substrate.

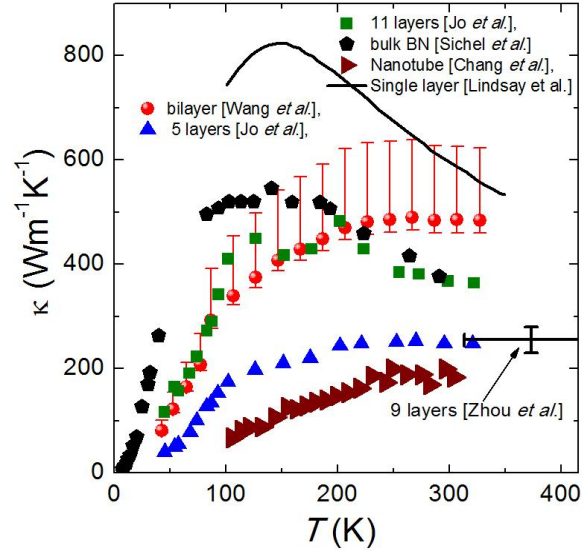

**Figure 6**. Experimental measurements of thermal conductivity in suspended few-layer *h*-BN. Data points for boron nitride nanotubes [97] and 9-layers *h*-BN [95] are shown for comparison. Reprinted from [98].

These existing experimental reports on thermal conductivity of few-layer *h*-BN are dominated by the organic residue on the surface which obscures the intrinsic phonon transport in *h*-BN. Therefore, brand-new transfer skill and measurement techniques should be invented to reveal the intrinsic phonon transport in *h*-BN and to examine the validation of anomalous size effect, dimensionality and anisotropy in this novel 2D system.

As such, Wang *et al*. fabricated suspended bilayer *h*-BN by dry-transfer method and reported that room temperature thermal conductivity is around 484 Wm-1K-1(+141 Wm-1K-1/ -24 Wm-1K-1) which exceeds that in bulk *h*-BN (red bullets in Figure 6) [98]. This PDMS-mediated dry-transfer method, whose sample quality, due to less polymer residues on surfaces, is believed to be superior to that of PMMA-mediated samples.

### 3.3 MoS$_2$

As a member of two-dimensional family, molybdenum disulfide (MoS$_2$) has a unique "sandwich" structure and natural thickness-depended energy gap [7, 99]. This unique property makes MoS$_2$ a promising candidate material for transistors instead of graphene. For materials to be used in high performance electronic or optoelectronic devices, high thermal conductivity and low interfacial thermal resistance is required to dissipate Joule heat efficiently and reduce temperature in the hot spot. To this end, several theoretical calculations have revealed the intrinsic thermal conduction of both monolayer and few-layer MoS$_2$ [37, 100-103], comparing to the rare experimental reports, leaving the intrinsic thermal conductivity of MoS$_2$ almost unclear [77, 104,

105].

**Table 2**. Room temperature thermal conductivity in few-layer MoS$_2$ from experiments.

| κ at RT | ~52 Wm$^{-1}$K$^{-1}$ | 34.5±4 Wm$^{-1}$K$^{-1}$ | 44-50 Wm$^{-1}$K$^{-1}$ | 48-52 Wm$^{-1}$K$^{-1}$ |
|---|---|---|---|---|
| Layers | 11 layers | 1 layer | 4 layers | 7 layers |
| samples | CVD | CVD | exfoliated | exfoliated |
| Method | 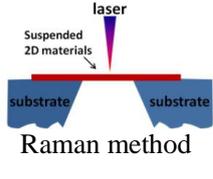 Raman method | 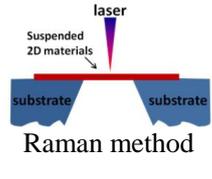 Raman method | 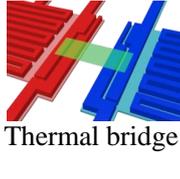 Thermal bridge | 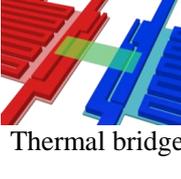 Thermal bridge |
| Geometry | Triangle | Corbino | Ribbon | Ribbon |
| Reference | [77] | [104] | [105] | [105] |

By combining the non-equilibrium Green's function approach and the first-principle method, Jiang *et al*. found that the room temperature thermal conductivity for the Armchair and Zigzag MoS$_2$ nanoribbons is about 674 Wm$^{-1}$K$^{-1}$ and 841 Wm$^{-1}$K$^{-1}$, respectively [106]. Cepellotti *et al*. claimed that the room temperature thermal conductivity of MoS$_2$ sheet reaches a value of around ~300 Wm$^{-1}$K$^{-1}$ using the exact solution of Boltzmann transport equation [89]. These theoretical calculations demonstrate that MoS$_2$ is a good thermal conductor whose thermal conductivity is much higher than that in silicon.

On the other hand, several other independent groups argued that MoS$_2$ possesses a thermal conductivity lower than that in silicon. Li *et al*. found that κ for a typical sample size of 1μm is 83 Wm$^{-1}$K$^{-1}$ using *ab* initio calculations [100]. Zhang *et al*. found the value is as low as around 26 Wm$^{-1}$K$^{-1}$ by phonon Boltzmann transport equation combined with relaxation time approximation [101]. Ding *et al*. found that the in-plane thermal conductivity of monolayer MoS$_2$ is about 20 Wm$^{-1}$K$^{-1}$ and reveals that the in-plane thermal conductivity of multilayer MoS$_2$ is insensitive to the number of layers due to the finite energy gap in the phonon spectrum of MoS$_2$, which makes the phonon–phonon scattering channel almost unchanged with increasing layer number [107]. Cai *et al*. found the room temperature around 23.2 Wm$^{-1}$K$^{-1}$ when solving the nonequilibrium Green's function [37]. The variations among different studies might be caused by the different approximations and force fields used in the calculations.

Despite the theoretical data scattered for more than one order of magnitude, the experimental results on few-layer MoS$_2$ seem to be more consistent with each other. Confocal micro-Raman method carried out by two independent group demonstrates that room temperature thermal conductivity in 1-layer and 11-layer MoS$_2$ is 34.5±4 Wm$^{-1}$K$^{-1}$ [104] and ~52 Wm$^{-1}$K$^{-1}$ [77], respectively (Table 2). Another group suspended few-layer MoS$_2$ on thermal bridge and obtained room-temperature thermal conductivity values to be (44–50) and (48–52) Wm$^{-1}$K$^{-1}$ for 4 and 7 layers, respectively [105]. These experimental observations are two to three folds smaller than that measured in MoS$_2$ thin flakes with 2μm to 5μm in thickness [108], which, again, is attributed to the scatterings from surface disorders, unfortunately.

### 3.4 Black phosphorous

Black phosphorous thin flakes was obtained by experimentalists very recently [6]. As the most stable allotrope of phosphorus at ambient condition, black phosphorous is a layered material with a direct band gap of 0.3 eV for bulk [109], which, combining its excellent thermal properties, is therefore proposed to be potential candidate for thermoelectric materials [110]. Thermal conductivity measurements were reported by three different groups using confocal micro-Raman method, thermal bridge method and TDTR, respectively [73, 111-113]. The measured value ranges from ~10 Wm$^{-1}$K$^{-1}$ to ~34±4 Wm$^{-1}$K$^{-1}$ in Zigzag direction and 17 Wm$^{-1}$K$^{-1}$ to 86±8 Wm$^{-1}$K$^{-1}$ in Armchair direction, depending on the thickness of flakes. For the anisotropic effect of black phosphorous, please refer to the detailed discussion in Sec.6.

### 3.5 Silicene

Silicene, the silicon counterpart of graphene, has been proposed to have better electronic properties and nontrivial phonon thermal conductivity. Unlike graphene in which all carbon atoms form honey-cone structures within a flat plane, silicon atoms in silicone show a buckled structure, resulting in unique thermal transports fundamentally differing from that in other 2D materials, namely (a) longitudinal and transverse acoustic phonons dominate the thermal transports and the acoustic out-of-plane phonon modes only have less than 10% contributions to the total thermal conductivity[114-117]; (b) thermal conductivity increases dramatically with tensile strain due to enhancement in acoustic phonon lifetime[118-120]. Unfortunately, no experiment on thermal conductivity in silicene has been reported.

### 4. Length dependence of phonon thermal transport

Thermal conductivity in 2D materials demonstrates anomalous size-dependence, comparing to the size-independence thermal conductivity in bulk materials. Theoretical studies on 2D Fermi-Pasta-Ulam (FPU) lattice predicted a logarithmical divergent thermal conductivity [24, 26, 28, 29]. In real 2D materials, such as graphene, Lindsay *et al*. theoretically found that the thermal conductivity in graphene is length dependent, due to selection rules for three phonon scattering and the phase space of which is strongly restricted by the reduced dimensionality [39]. Nika *et al*. emphasized the importance of low-frequency acoustic phonons in graphene and predicted that the thermal conductivity of graphene should be increased with sample length when $L < 30$ μm [38]. Zhu *et al*. predicted the coexistence of size-dependent and size-independent thermal conductivities for single layer black phosphorus (BP) along Zigzag and Armchair directions, respectively [121].

Xu *et al*. [11] carried out the first experimental on length-dependent thermal conductivity in suspended 2D materials and found thermal conductivity in suspended single layer graphene diverges with sample length as $\kappa \sim \log L$ (Figure 7e), which is due to the 2D nature of phonons in graphene, and is consistent with theories on FPU lattice and their own MD simulations. The experiment was carried out on CVD graphene using thermal bridge method (Figure 7 a-d). As mentioned above, directly measuring the thermal contact resistance $R_c$ is challenging in studying the thermal transport in graphene, the authors assumed $2R_c$ contribute 0% (red squares in Figure 7e), 5% (blue circles in Figure 7e) and 11.6% (wine pentagons in Figure 7e) to the total measured thermal resistance $R_{total}$, according to their assumptions and calculations.

In the same experiment, Xu *et al*. fixed sample width to 1.5 μm, as the two edges may also affect thermal conductivity by scattering the phonons, especially the phonons with long mean free path. Figure 7f shows the thermal conductivity changes from ~1054 Wm$^{-1}$K$^{-1}$ to ~1186 Wm$^{-1}$K$^{-1}$ when changing sample width from 1.5μm to 2.5μm. This observed weak width-dependent thermal conductivity can be explained by a simple empirical model [15]:

$$\kappa(w) \approx \left[ \frac{1}{c}\left(\frac{\Delta}{w}\right)^n + \frac{1}{\kappa} \right]^{-1}, \qquad (9)$$

where $\Delta$ is mean square root of roughness on the edge, $c$ and $n$ are fitting parameters. In Figure 7g, the parameter for $\Delta$, $c$ and $n$ is 0.6nm, 0.04 Wm$^{-1}$K$^{-1}$ and 1.8, respectively. Similar width-dependent was also observed in supported graphene [15].

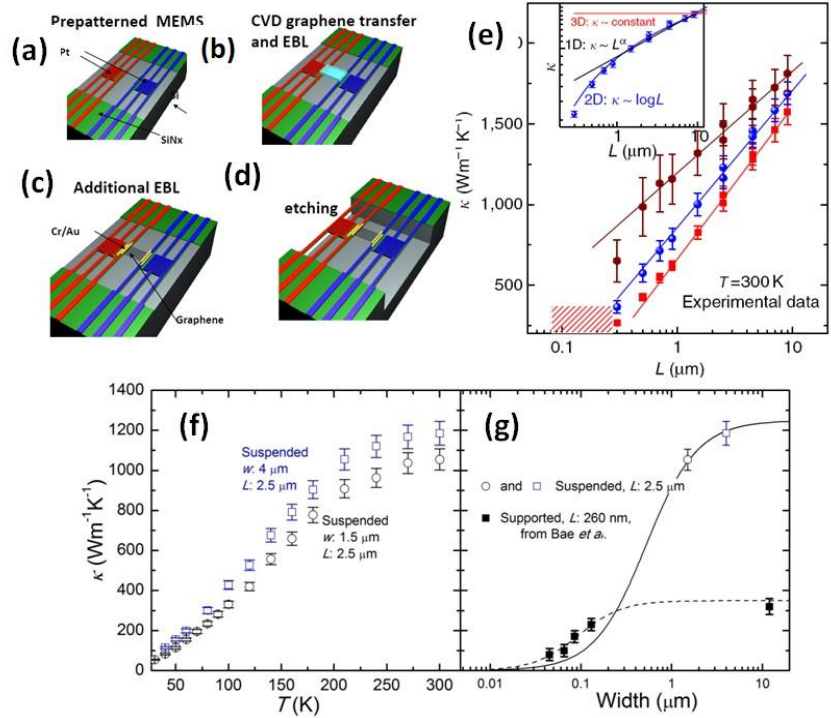

Figure 7. Length-dependent thermal conductivity measured in suspended single layer graphene. (a)-(d) Fabricating CVD graphene suitable for thermal bridge method. It is worth noting that the additional step of EBL for Cr/Au bars on two ends of graphene helps to improve the thermal contact resistance. (e) Observation of length-dependent thermal conductivity when thermal contact resistance $2R_c$ contributes 0% (red squares), 5% (blue circles) and 11.6% (wine pentagons) to the total measured thermal resistance $R_{total}$. (f) and (g) Width-dependent thermal conductivity, the data labeled by solid black squares are supported graphene from reference[15]. Reproduced with permission from [11]. Copyright 2014 Nature Publishing Group.

## 5. Thickness/number of layers dependence of phonon thermal transport

Heat conducts differently in nanostructures when shrinking from bulk into 2D structure. Of particular fundamental interest is the dimensional crossover from 3D into 2D. Ghosh *et al*. found that at room temperature thermal conductivity reaches as high as ~ 4000 $Wm^{-1}K^{-1}$ in suspended single layer graphene, and changes from ~2800 $Wm^{-1}K^{-1}$ to ~1300 $Wm^{-1}K^{-1}$ when the number of atomic plane in few layer graphene increase from 2 to 4 (red squares in Figure 8a) [12]. Authors attributed the observed results to the cross-plane coupling of low-frequency phonons and the enhancement of phonon scattering between layers.

On the other hands, experiments on supported graphene [122] and encased graphene (green circles in Figure 8a) demonstrate that thermal conductivity increases with graphene layers [123]. On the amorphous silicon dioxide substrate, both the theoretical [124] and experimental [122] studies show that thermal conductivity of supported multi-layer graphene saturates to that of bulk graphite at the thickness of about 40 layers. Similar thickness-dependence behavior also was observed by several independent groups on black phosphorous [73, 111, 112], and $MoS_2$ [108].

These two different trends with respect to thickness are understandable as the interaction between 2D materials and substrate materials can also enhance the phonon scatterings in the 2D materials layers. Furthermore, the polymer residues/roughness on graphene/MoS$_2$/h-BN/black phosphorous surfaces can also increase the phonon scatterings, resulting in opposite trend of the thickness dependent thermal conductivity with respect to that in suspended 2D materials (Figure 8b).

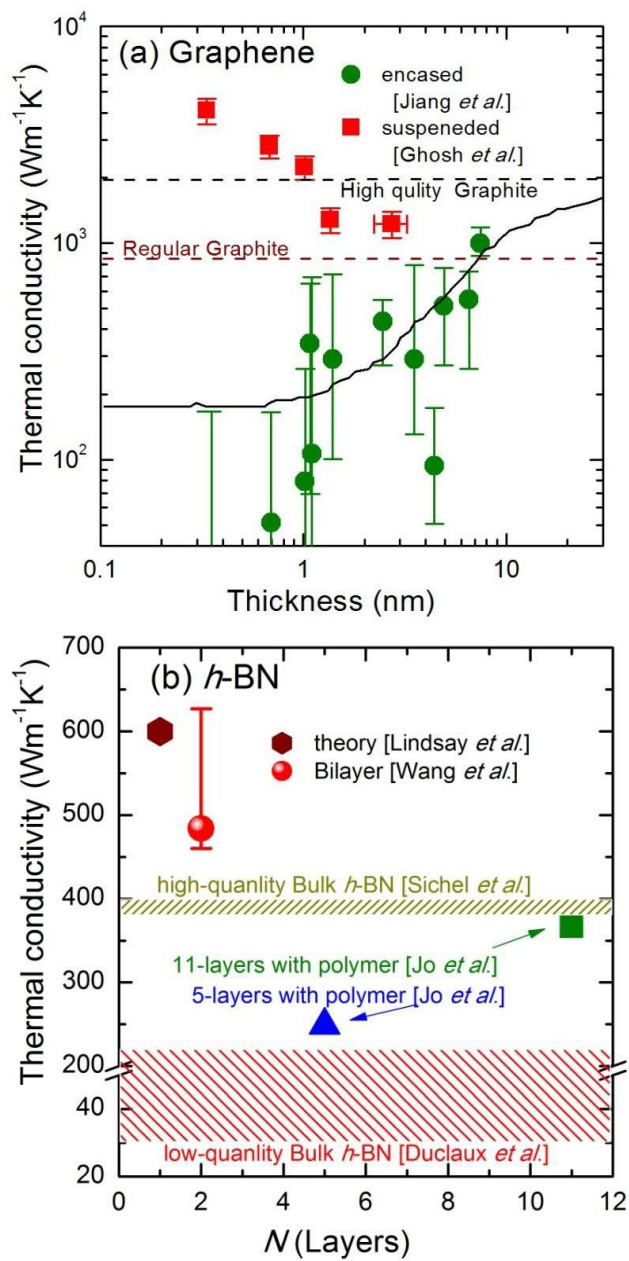

Figure 8. (a)Two different trends on thickness-dependent thermal conductivity of suspended and encased graphene, data adopted from [12]&[123]. (b) Layer-dependent thermal conductivity in h-BN, reprinted from [98].

## 6. Anisotropic effect of phonon thermal transport

Layered-structure materials, e.g. 2D materials, mica and high $T_c$ superconductors,

demonstrate strong anisotropic thermal conductivity between in-plane direction and out-of-plane direction, with the experimentally observed anisotropic ratio reaches as high as ~100 and ~300 in bulk *h*-BN and graphite [125]. More interestingly and very recently, theoretical approaches demonstrated that in-plane thermal conductivity in Zigzag and Armchair direction shows anisotropic behavior in black phosphorene, with the predicted value of 110 $Wm^{-1}K^{-1}$ and 36 $m^{-1}K^{-1}$ along Zigzag and Armchair direction, respectively [126]. This is due to the fact that phosphorous atoms present a Great-Wall like structure along Armchair direction, differing from the straight line along Zigzag direction, resulting in strong anisotropic behavior in thermal properties, electronic properties and optical properties in black phosphorous.

The room temperature Raman method results of anisotropic behavior in black phosphorous flakes shows the anisotropic ratio is between 1.5 and 2. The Armchair and Zigzag thermal conductivity are ~12 $Wm^{-1}K^{-1}$ and ~18 $Wm^{-1}K^{-1}$ (anisotropic ratio is ~1.5) for 9.5nm-thick film, and increase to ~20 $Wm^{-1}K^{-1}$ and ~40 $Wm^{-1}K^{-1}$ for 15nm-thick film (anisotropic ratio is ~2) [73]. This thickness dependent anisotropic effect also observed by Jang *et al*. using TDTR method and the measured thermal conductivity along the zigzag direction (~86 ± 8 $Wm^{-1}K^{-1}$) is ~2.5 times higher than that of the armchair direction (34 ± 4 $Wm^{-1}K^{-1}$) for black phosphorous flakes with thickness ranging from 138nm to 552nm [111].

Another experiment carried out by Lee *et al*. on black phosphorous ribbons demonstrates increasing thermal conductivity anisotropy of around two with temperature above 100K using thermal bridge method. They attributed this to orientation-dependent phonon dispersion and phonon-phonon scattering based on the density function perturbation theory [112].

**7. Basics of phonon thermal transport in 2D materials**

**7.1 Theoretical approaches**

Various approaches have been used to investigate the thermal transport in two-dimensional materials, including molecular dynamics (MD) simulations[32-34], non-equilibrium Green's function (NEGF) method[35-37], and Boltzmann transport equation (BTE)[38-40, 127, 128]. In this section, we briefly review these theoretical approaches.

**Molecular Dynamics Simulations**

MD simulation is a classical approach that can model the dynamics of each atom in a system of interest based on the Newton's equation of motion and the empirical force field. It has a number of advantages, such as the capability of modeling complex material systems with a large number of atoms, considering atomic level details including defects, strain, surface reconstruction, etc., and accounting for anharmonicity to all orders. In the non-equilibrium MD (NEMD) simulations, two thermostats [129] at different temperatures (Figure 9a) are used to mimic the heat source and sink in experiment. Thermal conductivity can be calculated from the Fourier's law of heat conduction

$$k = -\frac{J}{\nabla T}, \tag{10}$$

where $J$ is the heat flux defined as the energy transported per unit time across unit area, and $\nabla T$ is the temperature gradient along the heat transport direction. In two-dimensional materials, such as single-layer graphene (SLG), the cross sectional area is usually defined as $S=W*d$, where $W$ is the width of the sheet, and $d$ is the inter-layer distance in the bulk (d=3.35 Å for graphene). In the steady state, heat flux can be calculated according to the energy injected into /extracted from the heat source /sink, and the temperature gradient can be calculated from the linear fit of the temperature profile [34]. One needs to run the simulation sufficiently long to reach the steady state where the heat flux and temperature profile is constant.

Alternatively, no temperature gradient is required in equilibrium MD (EMD) simulation, and the thermal conductivity (tensor) can be computed from the Green-Kubo formula

$$k_{mn} = \frac{1}{k_B T^2 V} \int_0^\infty \langle J_m(0) J_n(t) \rangle dt, \tag{11}$$

where $k_B$ denotes the Boltzmann constant, $T$ and $V$, denotes, respectively, the temperature and volume of the system, $J_\mu$ denotes the heat current in the $\mu$ direction, and the angular bracket denotes the ensemble average. The exact definition of heat current and the integration scheme in Eq. (11) is given by Schelling *et al.*[130] The total simulation time should be long enough to allow for the proper decay of the heat current auto-correlation function, and one should repeat the EMD simulation with different initial conditions to suppress the fluctuation [131]. More details about MD approach for heat conduction can be found elsewhere[132].

MD simulations have been widely used to study the thermal transport in two-dimensional materials from various aspects, such as the length dependent thermal conductivity in suspended SLG [11], the thickness dependent thermal conductivity in suspended [133] and supported [124] few-layer graphene (FLG), the c-axis thermal conductivity in intrinsic [134, 135] and strained [34] graphene, interfacial thermal transport across graphene-water interface [136], the effect of vacancy [131] and isotopic [137] defect, thermal rectification effect [138, 139], thermal conductivity of monolayer molybdenum disulfide (MoS$_2$) [140] and graphene-carbon nanotube hybrid [141].

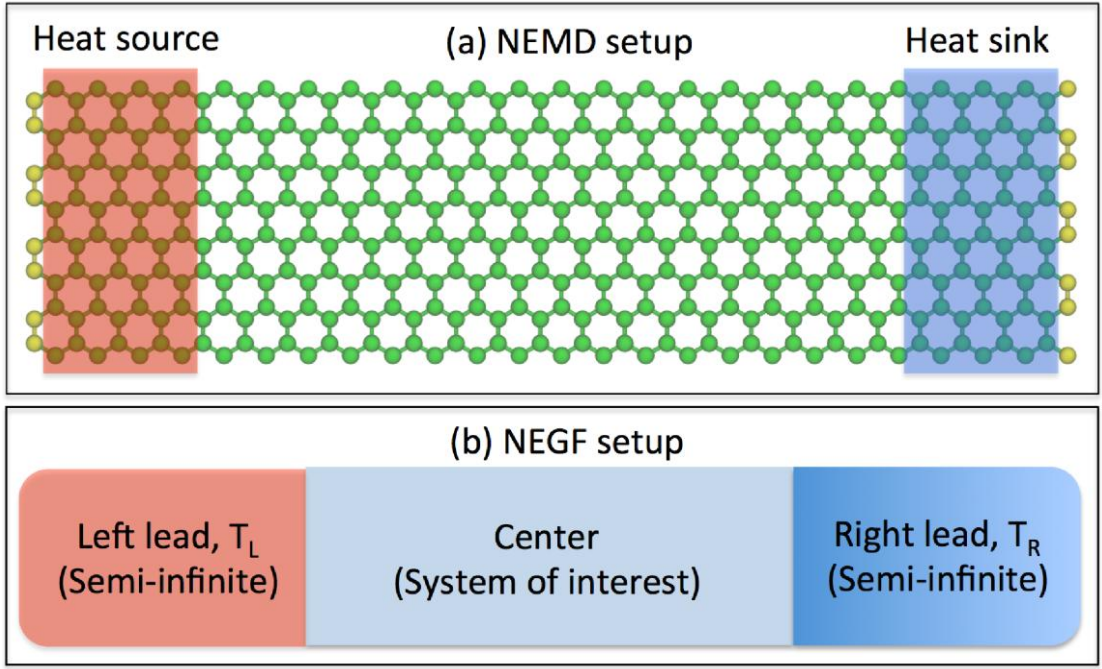

**Figure 9.** Schematic graph for the simulation setup. (a) NEMD setup. The yellow atoms are fixed boundary atoms. Two thermostats at different temperatures are used to impose the temperature gradient. (b) NEGF setup. The system of interest is sandwiched by two semi-infinite leads at different temperatures.

**Non-equilibrium Green's Function**

The non-equilibrium Green's function (NEGF) method, also known as atomistic Green's function method, is an elegant and powerful method to treat interacting systems at non-equilibrium in a rigorous way. It has its root from quantum field theory. NEGF was initially developed to handle electrical transport, but recently there have been a boost of applications of NEGF method in the study of thermal properties of two-dimensional materials [35-37, 142-149]. Readers are referred to the review articles by Zhang *et al.* [150] and Wang *et al.* [151, 152] for the details about NEGF methodology.

As shown in Figure 9b, the system of interest at the center is sandwiched by two semi-infinite leads at temperature $T_L$ and $T_R$, respectively. In the ballistic thermal transport limit, the heat current flowing from the left to the right lead is given by the Landauer formula

$$J = \int_0^\infty \frac{d\omega}{2\pi} \hbar\omega \zeta(\omega) \left[ f_L(\omega) - f_R(\omega) \right], \qquad (12)$$

where $\hbar$ is the Planck constant, $\omega$ is the phonon frequency, $\zeta(\omega)$ is the phonon transmission function, and $f_{L,R}(\omega) = \left\{ \exp\left[ \hbar\omega / (k_B T_{L,R}) \right] - 1 \right\}^{-1}$ is the Bose-Einstein distribution. In the limit of very small temperature difference between two leads, the thermal conductance can be written in a form similar to Landauer formula as

$$\sigma \equiv \lim_{\substack{T_L \to T \\ T_R \to T}} \frac{J}{T_L - T_R}$$
$$= \int_0^\infty \frac{d\omega}{2\pi} \hbar\omega \zeta(\omega) \frac{\partial f(\omega)}{\partial T} \quad . \tag{13}$$

In order to get the heat current or thermal conductance, the major task is to compute the phonon transmission function $\zeta(\omega)$. It was first shown by Caroli *et al.* [153] that the transmission function can be computed as (known as Caroli formula)

$$\zeta(\omega) = \text{Tr}\left[\Gamma_L(\omega) G^a(\omega) \Gamma_R(\omega) G^r(\omega)\right], \tag{14}$$

where Tr means taking the trace, $G^{r,a}(\omega)$ is the retarded or advanced Green's function for the central region related by $G^r(\omega) = \left(G^a(\omega)\right)^\dagger$, and $\Gamma_{L,R}(\omega)$ describes the coupling between the leads and the center.

If nonlinear interaction at the central part is further considered, the effective transmission function reads

$$\tilde{\zeta}(\omega) = \frac{1}{2}\text{Tr}\left[G^r\left(\Gamma_L + \frac{1}{2}\Gamma_n - S\right)G^a \Gamma_R\right]$$
$$+ \frac{1}{2}\text{Tr}\left[G^a \Gamma_L G^r\left(\Gamma_R + \frac{1}{2}\Gamma_n + S\right)\right], \tag{15}$$

where the nonlinear effect is reflected in the extra terms, $\Gamma_n = i\left(\Sigma_n^r - \Sigma_n^a\right)$ and

$$S = \left(f\frac{d\Gamma_n}{dT} - i\frac{d\Sigma_n^<}{dT}\right)\left[\frac{\partial f}{\partial T}\right]^{-1}, \tag{16}$$

Here the superscript < means the lesser Green's function, and $\Sigma_{L,R}^{r,a}$ is the retarded or advanced self-energy due to the coupling to the left or right lead. The detailed algorithms for the calculation of all the Green's functions are available in Ref. [151].

As a quantum approach, NEGF is particularly strong in handling low-temperature thermal transport, but has a limited capability of modeling systems with a large number of atoms. For simple system like graphene nanoribbons, NEGF method has been extensively used to study various effects on thermal transport, such as edge chirality[35], superlattice structure[142], structural and substitutional defects[36, 144, 147], hydrogen passivation[143], and strain[145]. Furthermore, NEGF method has also been used to handle more complex systems, such as the thermoelectric properties of graphene/boron nitride hybrid structure[146], the thermal conductivity of monolayer MoS$_2$[37], and the phonon and magnon Hall effect in two-dimensional lattice[148, 149].

**Boltzmann Transport Equation**

The phonon Boltzmann transport equation (BTE) was first formulated by Peierls[154] in 1929 as the microscopic description of the phonon heat conduction in dielectric crystals. Phonons in crystals can be characterized by the phonon wavevector **k**, and the polarization index *s*. For each phonon mode $\lambda(\mathbf{k},s)$, the phonon BTE describes in the steady state the balance between the phonon diffusion due to the temperature gradient and the phonon collision as

$$-\mathbf{v}_\lambda \cdot \nabla T \frac{\partial n_\lambda}{\partial T} + \left(\frac{\partial n_\lambda}{\partial t}\right)_{collision} = 0, \quad (17)$$

where **v** and *n* is the phonon group velocity and occupation number, respectively.

The major difficulty in solving phonon BTE comes from the collision term. The most common assumption for the collision term is the single-mode relaxation time approximation (SMRTA)

$$\left(\frac{\partial n_\lambda}{\partial t}\right)_{collision} = \frac{n_\lambda^0 - n_\lambda(t)}{\tau_\lambda}, \quad (18)$$

where $n_\lambda^0$ is the equilibrium phonon occupation number (Bose-Einstein distribution). It only accounts for the deviation from the equilibrium distribution for a single mode $\lambda$. Under this approximation, the lattice thermal conductivity can be computed as

$$\kappa_{ab} = \frac{1}{V} \sum_\lambda C_\lambda v_{\lambda a} v_{\lambda b} \tau_{\lambda b}, \quad (19)$$

where *V* is the volume, $C_\lambda = \partial E_\lambda / \partial T = k_B (\hbar \omega_\lambda / k_B T)^2 n_\lambda^0 (n_\lambda^0 + 1)$ is the specific heat per mode, and $v_{\lambda \alpha}$ is the phonon group velocity for mode $\lambda$ in the $\alpha$ (Cartesian) direction. Here $\sum_\lambda \to \sum_s \int d\mathbf{k}$ stands for the integration over the wavevectors for all branches. Take three-phonon scattering for instance, the phonon scattering processes satisfy the conservation of energy and quasi-momentum

$$\omega_\lambda \pm \omega_{\lambda'} = \omega_{\lambda''}, \quad \mathbf{k} \pm \mathbf{k'} = \mathbf{k''} + \mathbf{G}, \quad (20)$$

where $\omega$ is the phonon frequency, and **G** is the reciprocal lattice vector that is zero for normal (N) process and nonzero for umklapp (U) process. For 2D lattice such as graphene, additional selection rule[39] applies due to the reflection symmetry perpendicular to the 2D plane.

Various analytical expressions with fitting parameters have been proposed in early works to model the phonon relaxation time for the N process[155, 156], the U process[156-158], the boundary[159-161], the isotope and edge roughness [162], and the impurity [163] scatterings. Here we focus on the Klemens formulism for quasi-2D systems [85, 164] that has been recently[38, 86] used to study the thermal conductivity of graphene. Klemens[157] approximated the matrix element for three-phonon scattering in terms of the Grüneisen parameter. Under the long wavelength approximation (LWA) $k\sim\omega/v$ (only valid for small phonon wavevectors or linear phonon dispersion), Nika *et al.* [86] obtained the following phonon relaxation time for the U process based on the Klemens formulism

$$\frac{1}{\tau_U^{I,II}(\mathbf{k},s)} = \frac{\hbar \gamma_s^2(\mathbf{k})}{3\pi\rho v_s^2(\mathbf{k})} \sum_{s's''} \iint \omega_s(\mathbf{k})\omega_{s'}(\mathbf{k}')\omega_{s''}(\mathbf{k}'') \\ \times \left\{ n^0\left[\omega_{s'}(\mathbf{k}')\right] \mp n^0\left[\omega_{s''}(\mathbf{k}')\right] + \frac{1}{2} \mp \frac{1}{2} \right\} dk_\parallel' dk_\perp' , \quad (21)$$

where $g_s(\mathbf{k})$ is the mode-dependent Grüneisen parameter, $\rho$ is the mass density, and $k_\parallel'$ and $k_\perp'$ is the parallel and perpendicular component of the wavevector. The $\mp$ sign corresponds to the type-I ($w_l + w_{l'} = w_{l''}$) and type-II ($w_l = w_{l'} + w_{l''}$) three-phonon scatterings, respectively.

The fitting parameters in the empirical expression for the relaxation time require the benchmark with the existing experimental data, rendering it having limited predictive power, especially for new materials. Therefore, there are increasing recent studies[39, 40, 165-169] to rigorously evaluate the phonon relaxation time and thermal conductivity from the anharmonicity of interatomic forces, without any fitting parameter. Omini *et al.* [165] developed an iterative scheme to solve the linearized phonon BTE exactly without resorting to the relaxation time approximation. The phonon occupation number deviates from the equilibrium distribution in the presence of a small temperature gradient $\nabla T$, and can be approximated to the first order as $n_l \approx n_l^0 + n_l^1$, where $n_\lambda^1 = -\Phi_\lambda \frac{\partial n_\lambda^0}{\partial(\hbar\omega_\lambda)} = \frac{n_\lambda^0(1+n_\lambda^0)}{k_B T}\Phi_\lambda$ with $\mathsf{F}_l$ measuring the deviation from the equilibrium. When only considering three-phonon scatterings for the intrinsic thermal conductivity, the linearized phonon BTE reads[165]

$$k_B T \mathbf{v}_l \cdot \nabla T \frac{\partial n_l^0}{\partial T} = \sum_{l'l''} \left[ W_{ll'l''}^+ \left(\mathsf{F}_{l''} - \mathsf{F}_{l'} - \mathsf{F}_l\right) + \frac{1}{2}W_{ll'l''}^-\left(\mathsf{F}_{l''} + \mathsf{F}_{l'} - \mathsf{F}_l\right)\right], \quad (22)$$

Here $W_{ll'l''}^\pm$ denote the three-phonon scattering rate for the type-I and type-II scattering, respectively, which can be determined from the Fermi's golden rule [166] Additional terms can be added to the right hand side of Eq. (22) to account for the other scattering mechanisms such as impurity and boundary scattering [166].

By defining $\mathsf{F}_l = \sum_a F_{la} \nabla_a T$, the linearized phonon BTE can be simplified as

$$F_{la} = F_{la}^0 + \mathrm{D}F_{la}, \tag{23}$$

where $\alpha = x, y, z$ is the Cartesian component, and

$$F_{\lambda\alpha}^0 = \frac{\hbar \omega_\lambda n_\lambda^0 (n_\lambda^0 + 1) v_{\lambda\alpha}}{TP_\lambda}, \tag{24}$$

$$\mathrm{D}F_{la} = \frac{1}{P_l} \left\{ \sum_{l'l''} \left[ W_{ll'l''}^+ (F_{l''a} - F_{l'a}) + \frac{1}{2} W_{ll'l''}^- (F_{l''a} + F_{l'a}) \right] \right\}, \tag{25}$$

$$P_l = \sum_{l'l''} \left( W_{ll'l''}^+ + \frac{1}{2} W_{ll'l''}^- \right). \tag{26}$$

In this form, the linearized phonon BTE can be solved exactly via the iterative scheme[165]. Once the convergence is reached, based on the definition of heat current $\mathbf{J} = \sum_\lambda \hbar \omega_\lambda n_\lambda^1 \mathbf{v}_\lambda$ and Fourier's law of heat conduction $J_a = \sum_b k_{ab} \nabla_b T$, the thermal conductivity tensor can be computed as

$$\kappa_{\alpha\beta} = \frac{1}{k_B TV} \sum_\lambda \hbar \omega_\lambda v_{\lambda\alpha} n_\lambda^0 (n_\lambda^0 + 1) F_{\lambda\beta}. \tag{27}$$

Compared to SMRTA where only the deviation from equilibrium distribution for a single mode is considered (Eq. (18)), the iterative scheme also takes into account the non-equilibrium distribution of other phonons ($F_{l'a}$ and $F_{l''a}$ in Eq. (25)) that participate in the three-phonon scattering process. Furthermore, by setting $F_{l'a} = F_{l''a} = 0$, the zeroth-order solution of Eq. (23) is equivalent to the SMRTA, such that Eq. (27) can be reduced to Eq. (19) with

$$\tau_{\lambda\beta} = \frac{TF_{\lambda\alpha}}{\hbar \omega_\lambda v_{\lambda\beta}}. \tag{28}$$

As a perturbation approach, most of the iterative BTE calculations are limited to only consider the leading order anharmonic perturbation, i.e., three-phonon scattering. In contrast, the phonon relaxation time can also be computed from MD simulations where the anharmonicity is modeled explicitly to all orders. Based on the harmonic approximation and the normal mode coordinate

$$Q(\mathbf{k}, s, t) = \sum_{lb} \sum_\alpha \sqrt{\frac{m_b}{N}} e^{i\mathbf{k} \cdot \mathbf{R}_l} \varepsilon_{\alpha b}^*(\mathbf{k}, s) x_\alpha(lb, t), \tag{29}$$

where $x_\alpha(lb, t)$ is the $\alpha$ component of the displacement vector for the $b$th atom in the $l$th unit cell, the total energy of the system is

$$E_t(\mathbf{k},s,t) = \frac{\omega Q^* Q}{2} + \frac{\dot{Q}^* \dot{Q}}{2}. \tag{30}$$

By tracking the decay of the energy auto-correlation function in the time domain, the relaxation time for individual phonon mode can be obtained from[170]

$$\frac{\langle E_t(\mathbf{k},s,t)E_t(\mathbf{k},s,0)\rangle}{\langle E_t(\mathbf{k},s,0)E_t(\mathbf{k},s,0)\rangle} = e^{-t/\tau(\mathbf{k},s)}. \tag{31}$$

Alternatively, the relaxation time can also be determined from the frequency domain by fitting the spectral energy density (SED) according to the Lorentzian function as[171]

$$E_t(\mathbf{k},s,\omega) = \frac{C(\mathbf{k},s)\Gamma(\mathbf{k},s)}{[\omega_a(\mathbf{k},s)-\omega]^2 + \Gamma^2(\mathbf{k},s)}, \tag{32}$$

$$t = \frac{1}{2\Gamma}, \tag{33}$$

where $C(\mathbf{k},s)$ is the mode-dependent constant, and $\omega_a(\mathbf{k},s)$ is the anharmonic phonon frequency incorporating the frequency shift due to anharmonicity.

In addition to the k-space based approaches, several other numerical methods for solving phonon BTE exist in literature. Chen proposed the ballistic-diffusive heat conduction equation in which the phonon distribution function was divided into ballistic and diffusive parts [172, 173]. Finite volume method has been used to numerically solve BTE for the thermal transport modeling [174, 175]. The discrete coordinate method, which is based on a selection of a finite set of propagation directions and is widely used in the radiative heat transfer, has also been employed to resolve the phonon BTE [176]. By treating phonons as quasi-particles, the phonon BTE can be solved via the particle-based method such as lattice Boltzmann method (LBM). Nabovati *et al.* [177] employed LBM to model the phonon transport in 2D systems based on the *D2Q9* and *D2Q7* lattice. Furthermore, the stochastic Monte Carlo method [178] has also been used to solve phonon BTE. Mei *et al.* [179] shown in a recent study that the experimental thermal conductivity results for the pure and isotopically modified graphene samples could be accurately reproduced by using Monte Carlo method.

## 7.2 Mode contributions to thermal conductivity

One intriguing problem in the studies on the thermal transport properties of graphene is which phonon polarization may account for the superior high thermal conductivity of SLG. To address this problem, different theoretical approaches have been used in various studies[38, 86, 180-184], which are summarized in Table 3.

Theoretical studies[38, 86, 180] based on phonon BTE calculations provide contradictive views on importance of flexural phonons (out-of-plane vibrations) contribution to the thermal conductivity of graphene. Based on Klemens approximation for the three-phonon scattering and LWA, Nika *et al.* [38, 86] introduced separate U process limited phonon relaxation time for LA and TA phonons in terms of the mode-dependent Grüneisen parameter $\gamma_s$ and phonon group velocity $v_s$. In this formalism, as the relaxation time $\tau$ is proportional to $(v_s/\gamma_s)^2$, the contribution from flexural acoustic (ZA) phonon is simply neglected due to its large Grüneisen parameter and small group velocity, particularly for the long wavelength ZA phonon. Nika *et al.* [38, 86] found the U process limited thermal conductivity value for SLG depends sensitively on the choice of Grüneisen parameter used in the calculations. Using Grüneisen parameter obtained from *ab initio* calculations, they reported room temperature thermal conductivity ($\kappa_0$)~ 4000 W/m-K [86]. Furthermore, they found from their calculations the relative contribution to thermal conductivity is LA~71% & TA~28.5% at 100 K, and LA~50% & TA~49% at 400 K.

Other works included the flexural phonons into the calculations explicitly. Within the framework of Callaway's effective relaxation time theory, Alofi *et al.*[180] computed the thermal conductivity contribution for three acoustic branches. They assumed the relaxation time for the N process and U process having the same frequency and temperature dependence as[180]

$$\tau_{anh}^{-1} = \left[ B_N + B_U e^{-\Theta/\alpha T} \right] \omega^2 T^3, \qquad (34)$$

where $B_N$ and $B_U$ are fitting parameters for the three-phonon N process and U process, respectively, $\Theta$ is the averaged Debye temperature for all acoustic branches, and $\alpha$ is a constant. They found ZA phonon makes the dominant contribution to the total thermal conductivity for all the temperatures (0~1000 K). At room temperature, the total thermal conductivity of SLG is $\kappa_0$~ 4100 W/m-K, with relative contribution as ZA~50%, TA~27%, and LA~23%. They attributed the dominant contribution from ZA phonon to the larger relaxation time and specific heat for the ZA phonon compared to the LA and TA phonons.

Without resorting to the empirical expressions for the relaxation time, several groups[181-184] used molecular dynamics and lattice dynamics calculations to compute explicitly the phonon relaxation time from the SED analysis in EMD simulations, and obtain mode-wise thermal conductivity in the framework of SMRTA. Using optimized Tersoff potential[185], Qiu *et al.*[181, 182] computed the relaxation time for all six phonon branches in suspended SLG of size 4.4 nm × 4.3 nm. Compared to the in-plane phonons, the flexural phonons (both ZA and ZO) are found to have a longer relaxation time, in the range of 10-40 ps at room temperature. They further calculated thermal conductivity with the quantum phonon (Bose-Einstein) distribution, and found $\kappa_0$=1626 W/m-K at room temperature. The mode contribution is LA~26%, TA~32%, ZA~29%, and a non-negligible contribution from optical mode ZO~ 13%. Using the same force field, Chen et al.[183] computed mode-wise thermal conductivity from 6 nm × 6 nm suspended SLG. Using classical phonon distribution ($n = \frac{k_B T}{\hbar \omega}$), they obtained $\kappa_0$ around 1607 W/m-K, and mode contribution is ZA~22%, TA~21%, LA~41%, ZO~11%, and LO~5%. With quantum phonon distribution, $\kappa_0$ is 904 W/m-K, and the mode contribution is ZA~35%, TA~26%, LA~34%, ZO~4%, and LO~1%. Based on SED analysis, Wei et al.[184] computed the mode dependent thermal conductivity of suspended SLG with size 8.76 nm × 7.58 nm using the original Tersoff potential[186]. They found the use of classical phonon distribution overestimates the thermal conductivity at low temperature compared to the quantum phonon distribution, resulting in 58% discrepancy in room temperature thermal conductivity $\kappa_0$. With quantum distribution, they obtained $\kappa_0$~3300 W/m-K, and the relative contribution is ZA~27%, TA~24%, LA~35%, and ZO~14%. In addition, the relative contribution from flexural phonons (ZA+ZO) is found to decrease with temperature, from 41% at 300 K to 32% at 1000 K.

Regarding the mode contribution, it is worth pointing out the molecular dynamics study by Zhang et al.[131] that evaluated the importance of flexural phonons from the Green-Kubo method. Using the optimized REBO potential[185], they obtained $\kappa_0$~2900 W/m-K. They further estimated the mode contribution via the freezing method. In their study, atomic motions are restricted to in-plane vibration or out-of-plane vibration by freezing certain degree of freedom. Additional Green-Kubo calculations are performed for cases with the restricted vibrations, from which they found the flexural phonon contribution is around 43%. In this study, no free parameter or assumption such as SMRTA is used. The results from MD simulations are known to be dependent on the choice of the force field. By comparing these MD based studies[131, 181-184], one can see that although the exact value of the predicted thermal conductivity (e.g., $\kappa_0$) differs to some extent, the predictions for the relative contribution of flexural phonons (~40%) agree quite well among these studies.

Without using SMRTA, Lindsay et al.[39] computed the phonon scattering rate from the Fermi's golden rule and solved phonon BTE exactly via an iterative approach. The optimized Tersoff potential[185] is used in their calculations, and phonon-phonon anharmonic interactions are limited to the three-phonon scattering process. They obtained $\kappa_0$~3400 W/m-K for SLG with the length of 10 μm, and the relative contribution is ZA~75%, TA~16%, and LA~9% at room temperature. The anomalously large contribution from ZA phonon is attributed to the two unique features of graphene. Firstly, compared to the in-plane branches, ZA phonon in graphene is found to have much larger density of states at small wavevectors due to its quadratic dispersion. This finding is later confirmed by the MD simulation[131]. In addition, they found the reflection symmetry of graphene imposed additional selection rule for the three-phonon scattering events that only even numbers of flexural phonons can be involved. This selection rule has profound effect on the thermal transport in graphene. They found about 60% of both N process and U process in the phase space of ZA phonon is forbidden by the selection rule, leading to the greatly suppressed three-phonon scattering for ZA phonon. This is consistent with the Wei *et al.*'s results[184] from MD simulation that ZA phonon has very long lifetime approaching 100 ps at low frequency. Using the same method and force field as Lindsay *et al.*'s work, Singh *et al.* [40] computed mode dependent thermal conductivity of graphene, with a different boundary scattering term. They also found ZA phonon contributed most significantly to the total thermal conductivity. They reported room temperature thermal conductivity ~3200 W/m-K, with relative contribution ZA~89%, TA~8%, LA~3%. Furthermore, they found at small wavevectors, the relaxation time for ZA phonon is more than one order of magnitude higher than that of TA and LA phonons.

Lindsay *et al.* [39] questioned the validity of LWA ($k\sim\omega/v$) for the calculation of three-phonon scattering rate. Firstly, the N process is typically not included in the LWA models of phonon BTE calculations[38, 86]. Although the N process does not directly contribute to the thermal resistance, it should not be neglected due to its essential role of redistributing phonons to larger wavevectors so that U process can occur. Singh *et al.*[187] found in their BTE calculations that the neglecting N process can lead to a significant over-prediction of thermal conductivity, and the divergence in thermal conductivity with increasing size. Furthermore, the expression for the three-phonon scattering matrix element in the LWA is proportional to *kk'k''*, which implicitly assumes that all the wavevectors are small. However, this assumption is in principle not compatible with the U process that causes the thermal resistance. By comparing results from the rigorous calculations of three-phonon scattering rate based on Fermi's golden rule and that from the Klemens approximation, Singh *et al.*[187] critically evaluated the validity of LWA. They found thermal conductivity of graphene is greatly under-predicted when LWA is used, with the most significant discrepancy occurring for ZA phonon.

As the experimental measurement of the mode contribution is still quite challenging, people mainly resort to theoretical modeling to address this problem. However, as we already see in this section, all the theoretical approaches for modeling thermal transport have approximations and limitations. For example, the theoretical studies using selection rule for phonon scatterings only consider perfectly flat 2D plane with reflection symmetry. However, it is observed in experiment[188] that the actual sample of suspended SLG has nanoscale ripples that help to stabilize the structure. When measuring the thermal conductivity, the graphene may have contact with the heater and sensor. These factors would break the reflection symmetry, thus allowing ZA phonon to scattering. Furthermore, when treating the phonon-phonon scattering, most of the BTE calculations based on perturbation theory are limited to three-phonon scattering. Previous MD studies[181, 182] on SLG suggest that higher-order anharmonicity might not be negligible. To this end, it is still unclear how higher-order phonon-phonon scatterings would change the relative contributions to thermal conductivity from different polarizations. On the other hand, although the phonon lifetime can be computed from MD simulations which account for anharmonicity to all orders explicitly, the calculation of thermal conductivity relies on SMRTA, which treats N process as independent resistive process (see for instance Eq. (34)). Lindsay *et al.* [39] compared the thermal conductivity results ($\kappa_L$) from the exact solution to phonon BTE with that from SMRTA ($\kappa_{RTA}$). They found SMRTA considerably underestimates the thermal conductivity of SLG, and the ratio $\kappa_L/\kappa_{RTA}$ at room temperature for different polarizations is ZA~8, TA~3, and LA~2 at the length of 10 μm. Due to these complexities, the problem of mode contribution to thermal conductivity of SLG is still under debate, and further investigations are needed.

**Table 3**. Percentage contributions to thermal conductivity κ for acoustic phonons in single-layer graphene.

| $\kappa_{ZA}$ (%) | $\kappa_{LA}$ (%) | $\kappa_{TA}$ (%) | $\kappa_{total}$ (W/m-K) | Comments | References |
|---|---|---|---|---|---|
| 1 | 50 | 49 | ~2900 | Graphene size L=5 μm at T=400 K<br>BTE calculation based on Klemens formulism and long wavelength approximation<br>Umklapp scattering limited phonon lifetime<br>Mode-dependent Grüneisen parameter | Nika *et al.* [86] |
| 50 | 23 | 27 | ~4100 | Graphene size L=2.9 μm at T=300 K<br>BTE calculation based on Callaway's effective relaxation time theory<br>Consider both normal and Umklapp scattering | Alofi *et al.* [180] |
| 75 | 9 | 16 | ~3400 | Graphene size L=10 μm at T=300 K<br>Solve linearized BTE via the iterative process<br>Three-phonon scattering rate computed based on Fermi's golden rule and optimized Tersoff potential<br>Apply selection rule | Lindsay *et al.* [39] |
| 89 | 3 | 8 | ~3200 | Graphene size L>10 μm at T=300 K<br>Solve linearized BTE for a Corbino geometry<br>Three-phonon scattering rate computed based on Fermi's golden rule and optimized Tersoff potential<br>Apply selection rule | Singh *et al.* [40] |
| 29 | 26 | 32 | 1626 | Graphene size 4.4 nm × 4.3 nm at T=300 K<br>BTE calculation in the framework of SMRTA<br>Phonon lifetime computed from MD simulation with optimized Tersoff potential<br>Quantum phonon distribution | Qiu *et al.* [181, 182] |

| 22 | 41 | 21 | 1607 (Classical phonon distribution) | Graphene size 6 nm × 6 nm at T=300 K BTE calculation in the framework of SMRTA | Chen et al. [183] |
| --- | --- | --- | --- | --- | --- |
| 35 | 34 | 26 | 904 (Quantum phonon distribution) | Phonon lifetime computed from MD simulation with optimized Tersoff potential | |
| 27 | 35 | 24 | ~3300 | Graphene size 8.76 nm × 7.58 nm at T=300 K BTE calculation in the framework of SMRTA Phonon lifetime computed from MD simulation with original Tersoff potential Quantum phonon distribution | Wei et al.[184] |

## 8. Conclusions and Outlooks

In this review, we discussed the fundamental phonon thermal conduction in novel 2D materials, and outlined various experimental techniques and theoretical approaches to explain the nontrivial phonon thermal transport related to the 2D nature of phonons, e.g. length dependence, thickness dependence and anisotropic effect. Special attentions were given to the experimental observations, despite the hot debates ongoing.

Compared with one decade ago when the first member of 2D materials, graphene, was exfoliated, we have much clearer understanding on the fundamental phonon thermal conduction in 2D materials. However, there are still challenges and debates both experimentally and theoretically which deserve further investigations.

The primary challenge lies in the difficulties in measuring the thermal contact resistance and the detecting temperature distribution in micro/nano scale with high sensitivity, although recent advances and modified experimental techniques claimed to have partially solved these problems [79, 80]. The second problem is related to the physics behind the novel thermal conduction including mode contributions, electron-phonon coupling on the contact, phonon-phonon scatterings in 2D sheet with size approach few hundred micrometers and above [189, 190]. The third challenge related to fabrication process for suspending clean 2D materials as the residues on the surfaces obscure the intrinsic phonon thermal conduction in 2D materials.

**Acknowledgments**

This work was supported in part by National Natural Science Foundation of China (No. 11304227 & No. 11334007) and by the Fundamental Research Funds for the Central Universities (No. 2013KJ024), and J. C. was supported by the National Youth 1000 Talents Program in China and National Natural Science Foundation of China (No. 51506153).